\documentclass[10pt, journal, lettersize]{IEEEtran}
\IEEEoverridecommandlockouts
\usepackage{xspace,amsmath,amssymb,amsfonts,epsfig,subfigure,syntonly}
\usepackage{lipsum}
\usepackage{cite,bm,color,url,textcomp,empheq,boxedminipage}
\usepackage{algorithmicx}
\usepackage{indentfirst}
\usepackage{epstopdf,makecell}
\usepackage{empheq}
\usepackage{pifont}
\usepackage{stfloats}
\usepackage[noend]{algpseudocode}
\usepackage{tcolorbox}
\usepackage[linesnumbered,ruled]{algorithm2e}
\usepackage{graphicx,graphics}  
\usepackage{multirow,multicol}
\usepackage{psfrag}    
\usepackage{stfloats}
\usepackage{url}
\usepackage{lipsum}

\newcommand\blfootnote[1]{%
  \begingroup
  \renewcommand\thefootnote{}\footnote{#1}%
  \addtocounter{footnote}{-1}%
  \endgroup
}

\long\def\symbolfootnote[#1]#2{\begingroup
\def\thefootnote{\fnsymbol{footnote}}
\footnote[#1]{#2}\endgroup}
\psfull

\allowdisplaybreaks[4]

\title{\LARGE Large Language Model Based Multi-Objective Optimization for Integrated Sensing and Communications in UAV Networks}
\author{\IEEEauthorblockN{Haoyun Li, Ming Xiao, Kezhi Wang, Dong In Kim, and Merouane Debbah}}
\date{May 2024}

\begin{document}

\maketitle
\begin{abstract}
    This letter investigates an unmanned aerial vehicle (UAV) network with integrated sensing and communication (ISAC) systems, where multiple UAVs simultaneously sense the locations of ground users and provide communication services with radars. To find the trade-off between communication and sensing (C\&S) in the system, we formulate a multi-objective optimization problem (MOP) to maximize the total network utility and the localization Cramér-Rao bounds (CRB) of ground users, which jointly optimizes the deployment and power control of UAVs. Inspired by the huge potential of large language models (LLM) for prediction and inference, we propose an LLM-enabled decomposition-based multi-objective evolutionary algorithm (LEDMA) for solving the highly non-convex MOP. We first adopt a decomposition-based scheme to decompose the MOP into a series of optimization sub-problems. We second integrate LLMs as black-box search operators with MOP-specifically designed prompt engineering into the framework of MOEA to solve optimization sub-problems simultaneously. Numerical results demonstrate that the proposed LEDMA can find the clear trade-off between C\&S and outperforms baseline MOEAs in terms of obtained Pareto fronts and convergence.
\end{abstract}

\blfootnote{This work is supported in part by Horizon Europe COVER project under grant number 101086228, and UKRI grant EP/Y028031/1. The work of Kezhi is partially supported by the Royal Society Industry Fellow scheme under IF-R2-23200104.}
\blfootnote{H. Li and M. Xiao are with the Division of Information Science and Engineering, KTH Royal Institute of Technology, Stockholm 10044, Sweden (email: {haoyunl, mingx}@kth.se). K. Wang is with the Department of Computer Science, Brunel University
London, Uxbridge, Middlesex, UB8 3PH (email: kezhi.wang@brunel.ac.uk). Dong In Kim is with the Department
of Electrical and Computer Engineering, Sungkyunkwan University, Suwon
16419, South Korea (e-mail:dongin@skku.edu). M. Debbah is with the Department of Electrical Engineering and Computer
Science and the KU 6G Center, Khalifa University, Abu Dhabi 127788, UAE,
and also with CentraleSupelec, University Paris-Saclay, 91192 Gif-sur-Yvette,
France (E-mail: Merouane.Debbah@ku.ac.ae).}
\section{Introduction}
Beyond the fifth generation (B5G) such as the sixth generation (6G) mobile networks have envisioned future mobile networks not only to provide ubiquitous communication services but also to support high-precision sensing capabilities. Toward this end, jointly designing communication and sensing (C\&S) has motivated significant research interest and applications in integrated sensing and communication (ISAC) \cite{isac0}. 
Thanks to high-quality line-of-sight (LoS) links for air-to-ground (A2G) communications and the controllable mobility and agility of unmanned aerial vehicles (UAV), the UAV networks have been expected to provide wider coverage and enhance C\&S performance in ISAC systems \cite{UAVtuto}.

Since communications and sensing share the same signals, there is an inevitable trade-off between C\&S. To explore the non-negligible trade-off in the UAV networks, a multi-objective optimization problem (MOP) could be applied. \textcolor{black}{To solve the resulting MOP, the bio-inspired multi-objective evolutionary algorithm (MOEA) including the multi-objective evolutionary algorithm based on decomposition (MOEAD) \cite{MOEAD}, the non-dominated sorting genetic algorithm (NSGAII) \cite{mopsurvey2}, the reference vector guided evolutionary algorithm (RVEA) \cite{mopsurvey2}, the adaptive geometry estimation based MOEA (AGE-MOEA) \cite{mopsurvey2}, and the multi-objective differential evolution algorithm (MODEA) \cite{MODEA}, is considered as a promising approach to finding Pareto-optimal solutions even for the non-convex Pareto front. To be more specific, MOEAD breaks a multi-objective problem into smaller subproblems and solves them together. NSGAII sorts solutions by dominance and keeps diversity through crowding distance. RVEA uses reference vectors to guide the search toward diverse solutions, while AGE-MOEA focuses on improving the Pareto front shape. MODEA applies differential evolution to efficiently explore and optimize solutions in multi-objective problems. However, the complexity of MOEAs is still high in general. Recently, large language models (LLM) have demonstrated remarkable capabilities in reasoning and prediction, which inspire researchers to explore the potential of LLMs via integrating LLMs with evolutionary algorithms (EA) \cite{LLM1}, \cite{LLM3}.} 
However, all the above works only considered a simple multi-objective test suite with clear Pareto fronts to verify the effectiveness of integrating LLMs with EAs. 
It is significantly appealing to integrate LLMs with MOEAs for solving real-world engineering problems such as those in UAV-enabled ISAC networks. \textcolor{black}{Integrating LLMs into UAV-enabled ISAC networks enhances UAVs' ability to process real-time communication and sensory data, enabling faster responses to environmental changes and task demands, such as deployment, route planning, and resource management. Additionally, LLMs simplify UAV control through natural language, improving collaboration between UAVs and with users, ground stations, and cloud servers.}

Motivated by the above background, we seek to find the trade-off between C\&S in a multi-UAV-enabled ISAC system, in which ground base stations (GBS) are either destroyed or out of function for various reasons, e.g., disasters or damages. Then multiple UAVs are deployed as aerial BSs to sense the locations of multiple ground users cooperatively and provide communication services simultaneously. With the setup, our objective is to simultaneously maximize the total network utility of all users and minimize the Cramér-Rao bounds (CRB) of user locations by jointly optimizing the UAV deployment and transmission power controls. 
Inspired by the huge potential of LLMs, we propose an LLM-enabled MOEA to solve the MOP. Specifically, we first choose MOEAD \cite{MOEAD} as the main MOEA framework to decompose the original MOP into a number of optimization subproblems. We second integrate the LLM as a black-box search operator into the framework of the MOEA by specifically designing the prompt engineering, considering both the multi-objective functions and constraint satisfaction. To the best of our knowledge, we are the first to attempt to
apply LLMs in the optimization of multi-UAV-enabled ISAC. Numerical results demonstrate that our proposed LLM-enabled MOEA significantly outperforms baseline algorithms in terms of finding Pareto fronts and convergence. 

{\it{Notations}}: ${(\cdot)}^{\dagger}$ denotes the transpose, $a \propto b$ denotes $a$ is proportional to $b$, and $x\sim \mathcal{C} \mathcal{N}\left(\mu, \sigma^2\right)$ denotes that $x$ follows the circularly symmetric complex Gaussian distribution with mean $\mu$ and variance $\sigma^2$.
\section{System Model and Problem Formulation}
\subsection{System Model}
We consider a multi-UAV network-enabled ISAC system that consists of $K$ UAVs equipped with ISAC units and $M$ ground users. 
In the system, the multi-UAV network transmits ISAC signals for downlink data via frequency division multiple access (FDMA) to all ground users and simultaneously performs radar sensing to locate the ground users. Then the network decides the deployment and power control of each UAV. For illustration, let $\mathcal{K}=\{1,2, \dots, K\}$ denote the set of UAVs. Assume each UAV $k \in \mathcal{K}$ is located at $(x_k, y_k, H)$ in a three-dimensional (3D) coordinate system, where $H \geq 0$ denotes the altitude for UAVs, and $\mathbf{q}_{k} = [x_k, y_k]^{\dagger}$ denotes the horizontal location of UAV $k$. To simplify the illustration, we assume all UAVs have the same  $H$. Let $\mathcal{M}=\{1,2, \dots, M\}$ denote the set of ground users and assume each user $m$ is an extended target with its center of mass located at $\mathbf{w}_m=[u_m, v_m]^{\dagger}$ on the ground. During each time slot $t$, the multi-UAV network transmits a set of unit-power ISAC waveforms $\mathcal{S}=\{s_k(t)\}.$ For each UAV $k$, the transmitted ISAC waveform $s_k(t)$ is the combination of radar waveform $s^{\text{rad}}_k(t)$ and communication waveform $s^{\text{com}}_k(t)$, i.e., $s_k(t) = s^{\text{rad}}_k(t) + s^{\text{com}}_k(t)$, where $s^{\text{rad}}_k(t)$ is assumed to be orthogonal to all $s^{\text{com}}_{k'}(t), ~k'\in\mathcal{K}$. All radar waveforms are assumed to be orthogonal to each other and known to all UAVs, and all communication waveforms are uncorrelated with each other \cite{ISAC1}. 

According to \cite{godrich}, the spatially distributed multiple antennas among all UAVs can act as a distributed multi-input and multi-output (MIMO) radar system. Let $\tau_{k, m, j}=\frac{R_{k, m}+R_{j, m}}{c}$ denote the propagation delay from transmitter UAV $k$, reflected by ground user target $m$, and received by UAV $j$,
where $c$ denotes the speed of light, and $R_{k, m}, R_{j, m}$ denote respectively the distances from UAV $k$ and UAV $j$ to ground user $m$, i.e, $R_{k, m}=\sqrt{\left(x_k-u_m\right)^2+\left(y_k-v_m\right)^2 + H^2}$.
The radar echo signal received at UAV $j$, reflected by ground user $m$, transmitted from UAV $k$ can be expressed as 
\begin{equation}
    s^{\text{rad}}_{k, m, j}(t)=\sqrt{\alpha_{k, m, j} p^{\text{rad}}_{k}} l_{k, m, j} s^{\text{rad}}_k\left(t-\tau_{k, m, j}\right)+w^{\text{rad}}_{k, m, j}(t),
\end{equation}
where $p^{\text{rad}}_{k}$ denotes the radar sensing power of UAV $k$, $\alpha_{k, m, j} \propto \frac{1}{R^2_{k, m}R^2_{j, m}}$ represents the variation in the signal strength due to path loss effects,  $l_{k, m, j}$ is the target radar cross section (RCS), and $w^{\text{rad}}_{k, m, j}(t) \sim \mathcal{C} \mathcal{N}\left(0, \sigma_w^2\right)$. 

The CRB is chosen as the metric to evaluate the radar sensing performance of the distributed MIMO radar system, 
which can be obtained by taking the inverse of the Fisher Information matrix (FIM).  Following a series of matrix manipulations, we can obtain the lower bound of localization of user $m$ based on \cite{godrich} as 
\vspace{-5pt}
\begin{equation}
    \text{tr}\left(\mathbf{C}^{u,v}_m\left(\{\mathbf{q}_k\}, \mathbf{p}^\text{rad}\right)\right)=\frac{\mathbf{a}_m^{\dagger} \mathbf{p}^\text{rad}}{{\mathbf{p}^\text{rad}}^{\dagger} \mathbf{Q}_m \mathbf{p}^\text{rad}},
\end{equation}
where $\mathbf{p^{\text{rad}}}=[p^{\text{rad}}_1, \cdots, p^{\text{rad}}_K]^{\dagger}$, $\mathbf{a}_m=(\mathbf{b}_{a_m}+\mathbf{b}_{b_m})$, $\mathbf{Q}_m=\mathbf{b}_{a_m}\mathbf{b}_{b_m}^{\dagger}-\mathbf{b}_{c_m}\mathbf{b}_{c_m}^{\dagger}$, $\mathbf{b}_{a_m}=[b_{a_{1,m}}, \cdots, b_{a_{K,m}}]^{\dagger}$, $\mathbf{b}_{b_m}=[b_{b_{1,m}}, \cdots, b_{b_{K,m}}]^{\dagger}$, $\mathbf{b}_{c_m}=[b_{c_{1,m}}, \cdots, b_{c_{K,m}}]^{\dagger}$, $\xi=\frac{8 \pi^2 B^2}{\sigma_w^2 c^2}$ and
\vspace{-2pt}
\begin{equation}
b_{a_{k,m}}=\xi \sum_{j=1}^K\alpha_{k, m, j}\left|l_{k,m,j}\right|^2\left(\frac{x_k-u_m}{R_{k,m}}+\frac{x_j-u_m}{R_{j,m}}\right)^2, 
\end{equation}
\begin{equation}
b_{b_{k,m}}=\xi \sum_{j=1}^K\alpha_{k, m, j}\left|l_{k,m,j}\right|^2\left(\frac{y_k-v_m}{R_{k,m}}+\frac{y_j-v_m}{R_{j,m}}\right)^2,
\end{equation}
\vspace{-7pt}
\begin{align}
b_{c_{k,m}}&=\xi \sum_{j=1}^K\alpha_{k, m, j}\left|l_{k,m,j}\right|^2\left(\frac{x_k-u_m}{R_{k,m}}+\frac{x_j-u_m}{R_{j,m}}\right)\notag\\
&\times\left(\frac{y_k-v_m}{R_{k,m}}+\frac{y_j-v_m}{R_{j,m}}\right).
\end{align}

We assume that the channel state information (CSI) between UAVs and ground users can be obtained by channel estimation techniques. For simplicity, the communication links between UAVs and users are assumed to be dominated by LoS links. Therefore, the A2G channel from UAV $k$ to user $m$ follows the free-space path loss model, and the channel power gain is given by
\begin{equation}
    h_{k,m} = \sqrt{\rho_0 R^{-2}_{k,m}},
\end{equation}
where $\rho_0$ represents the channel power at the reference distance 1 m, and $R_{k,m}$ is the distance from UAV $k$ to ground user $m$. It is assumed that the signals reflected by other users at user $m$ have a significantly decreased magnitude compared with the LoS transmission from UAV $k$. 
Note that $\mathcal{S}^{\text{rad}}$ is orthogonal to $\mathcal{S}^{\text{com}}$ and assumed to be known to all UAVs. Thus, the signal-to-interference-plus-noise ratio (SINR) of the signal received by user $m$ from UAV $k$ is then given by
\begin{equation}
    \gamma_{k,m}\left(\{\mathbf{q}_k\}, \mathbf{p}^\text{com}\right)=\frac{p^{\text{com}}_k h^2_{k,m}}{\sum_{k^{\prime} \in \mathcal{K} \backslash k} p^{\text{com}}_{k^{\prime}} h_{k^{\prime}, m}^2+\sigma_w^2},
\end{equation}
where $\mathbf{p^{\text{com}}}=[p^{\text{com}}_1, \cdots, p^{\text{com}}_K]^{\dagger}$ denotes the communication power of all UAVs.
Each UAV is allocated a bandwidth of $B$, and each user served by the same UAV is assumed to be allocated with equal bandwidth. Thus, the achievable data transmission rate from UAV $k$ to user $m$ is
\begin{equation}
    r_{k, m}\left(\{\mathbf{q}_k\}, \mathbf{p}^\text{com}\right)=\frac{B}{M} \log _2\left(1+\gamma_{k, m}\left(\{\mathbf{q}_k\}, \mathbf{p}^\text{com}\right)\right).
\end{equation}
\subsection{Problem Formulation}
Our goal is to find the trade-off between C\&S in the multi-UAV network ISAC system. To evaluate the communication performance, we adopt a proportionally fair network utility optimization framework of maximizing the sum log-utility across all the users \cite{utility}, which is expressed as
\begin{equation}
    \mathcal{F}_1(\{\mathbf{q}_k\}, \mathbf{p}^\text{com}) = \sum\limits_{m \in \mathcal{M}} \text{log} \left(\sum\limits_{k \in \mathcal{K}} r_{k, m}\left(\{\mathbf{q}_k\}, \mathbf{p}^\text{com}\right)\right).
\end{equation}
To evaluate the performance of location sensing performance, we use the log sum of the CRBs of all ground users as the second object to be optimized, which is expressed as
\begin{equation}
    \mathcal{F}_2(\{\mathbf{q}_k\}, \mathbf{p}^\text{rad}) = \text{log}\left(\sum\limits_{m \in \mathcal{M}} \text{tr}\left(\mathbf{C}^{u,v}_m\left(\{\mathbf{q}_k\}, \mathbf{p}^\text{rad}\right)\right)\right).
\end{equation}

Based on the settings mentioned above, our goal is to maximize the total network utility while minimizing the log sum of CRBs by jointly optimizing $\{\mathbf{q}_k\}$, ${\{\mathbf{p}^{\text{rad}}_k\}}$, and $\{\mathbf{p}^{\text{com}}_k\}, \forall k \in \mathcal{K}$. Therefore, the optimization problem is a CMOP formulated as
\begin{subequations}
    \begin{align}
\text{(P1)}: &\min\limits_{\{\mathbf{q}_{k}\}, {\{\mathbf{p}^{\text{rad}}_k\}}, \{\mathbf{p}^{\text{com}}_k\}} [-\mathcal{F}_1, \mathcal{F}_2]\notag\\
     &\text {s.t.}\quad p^{\text{rad}}_k \geq 0, \forall k \in \mathcal{K},\label{a}\\
    &p^{\text{com}}_k \geq 0, \forall k \in \mathcal{K},\label{b}\\
    &p_{\text{min}} \leq p^{\text{rad}}_k+p^{\text{com}}_k \leq p_{\text{max}},\forall k \in \mathcal{K}, \label{c}\\
    &\mathbf{q}_{k} \in \mathcal{A}, \forall k \in \mathcal{K} \label{e},
\end{align}
\label{MOP}
\end{subequations}
where \eqref{a}, \eqref{b} respectively denote the radar sensing power and communication power of each UAV, which should be no less than zero, \eqref{c} denotes the transmission power of each UAV, between $p_{\text{min}}$ and $p_{\text{max}}$, and \eqref{e} denotes the location constraints of UAVs, which are limited in the area of interest, i.e., $\mathcal{A}$. 

It is clear that there is a trade-off between the two objects $\mathcal{F}_1$ and $\mathcal{F}_2$ in CMOP (P). 
\textcolor{black}{Note that our proposed CMOP is highly non-convex and NP-hard due to the non-convex objective functions. Thus, it is challenging to solve the problem and very hard to acquire the optimal closed-form solution directly. Therefore, in the next section, we propose an LLM-enabled decomposition-based MOEA (LEDMA) to solve our CMOP and acquire near-optimal solutions.}

\section{Proposed Algorithm}
To solve the problem, we first introduce the decomposition method of multi-objective optimization. Then we will introduce the overall framework of LEDMA in detail. For simplicity, we rewrite our proposed CMOP as
\begin{align}
    \text{(P2)}:~&\min\limits_{\{\mathbf{x}\}}~{\boldsymbol{\Tilde{\mathcal{F}}}(\mathbf{x})=[\Tilde{{\mathcal{F}}}_1(\mathbf{x}), \Tilde{\mathcal{F}}_2(\mathbf{x})]}\notag\\
    &~~\text {s.t.}~\mathbf{x} \in \boldsymbol{\mathcal{C}} \label{p},
\end{align}
where $\mathbf{x}=[\mathbf{q}^{\dagger}_{1}, \cdots, \mathbf{q}^{\dagger}_{K}, p^{\text{rad}}_1, \cdots, p^{\text{rad}}_K, p^{\text{com}}_1, \cdots, p^{\text{com}}_K]^{\dagger}$ represents all optimization variables, $\boldsymbol{\mathcal{C}}$ denotes the decision space, and $\Tilde{{\mathcal{F}}}_1(\mathbf{x})=-\mathcal{F}_1(\mathbf{x}),~\Tilde{\mathcal{F}}_2(\mathbf{x})=\mathcal{F}_2(\mathbf{x})$.
The decomposition of CMOP involves decomposing (P2) into a series of single-objective sub-problems based on the Tchebycheff method, in which the sub-problem is expressed as 
\begin{align}
    &\min\limits_{\{\mathbf{x}\}}~{f(\mathbf{x}|\boldsymbol{\omega}, \mathbf{z}^*)=\max\limits_{1 \leq i \leq 2}\{\omega_i(\Tilde{\mathcal{F}}_i(\mathbf{x}) - z_i^*)\}}\notag\\
    &~~\text {s.t.}~\mathbf{x} \in \boldsymbol{\mathcal{C}} \label{subp},
\end{align}
where $\boldsymbol{\omega}=[\omega_1, \omega_2]^{\dagger}$ denotes the weight vector satisfying $0 < \omega_1, \omega_2 < 1,~\omega_1+\omega_2=1$, and $\mathbf{z^*}=[z^*_1, z^*_2]^{\dagger}$ is the global reference point contains the current minimum objective function values, which is defined as $z_i^*=\min\limits\{\Tilde{\mathcal{F}}_i(\mathbf{x})\}$ for $i=1,2$. For Tchebycheff method, there always exists a weight vector $\boldsymbol{\omega^*}$ for each Pareto-optimal point $\mathbf{x}^*$ such that $\mathbf{x}^*$ is the optimal solution of \eqref{subp}, and each optimal solution of \eqref{subp} is a Pareto-optimal solution of \eqref{p} \cite{MOEAD}.

Inspired by \cite{LLM1, LLM3}, we integrate the LLM as the search operator into the framework of MOEAD \cite{MOEAD}, which consists of initialization, evolution, and update processes. In each iteration, we cooperatively solve all sub-problems and maintain an external population (EP) $\boldsymbol{\Psi}$ containing non-dominated solution points at the current iteration. The steps are as follows.
\subsubsection{Initialization}
The proposed LEDMA starts with initialized weight vectors with a population size of $N$, i.e., $\boldsymbol{\omega}^j, ~j=1,\cdots, N$. Furthermore, the original CMOP \eqref{p} is decomposed into $N$ single-objective sub-problem based on \eqref{subp}. The objective function to be minimized in the $j$-th sub-problem associated with $\boldsymbol{\omega}^j$ can be expressed as
\begin{equation}
    {f^j(\mathbf{x}|\boldsymbol{\omega}^j, \mathbf{z}^*)=\max\limits_{1 \leq i \leq 2}\{\omega^j_i(\Tilde{\mathcal{F}}_i(\mathbf{x}) - z_i^*)\}}.
\end{equation}
Accordingly, an initial population of size $N$ is uniformly and randomly generated from $\boldsymbol{\mathcal{C}}$, i.e., $\mathbf{x}^j \in \boldsymbol{\mathcal{C}},~j=1,\cdots,N$, where $\mathbf{x}^j$ represents the initial solution of the $j$-th sub-problem. Note that $f^j(\mathbf{x}|\boldsymbol{\omega}^j, \mathbf{z}^*)$ is continuous functions of $\boldsymbol{\omega}^j$, which indicates that the optimal solution of the $j$-th sub-problem is supposed to be close to that of the $j'$-th sub-problem if $\boldsymbol{\omega}^j$ is close enough to $\boldsymbol{\omega}^{j'}$ in terms of Euclidean distances. Hence, we define the neighbor set of $\boldsymbol{\omega}^j$ as $S$ indexes of weight vectors closest to $\boldsymbol{\omega}^j$, i.e., $\mathcal{N}(j)=\{j_1, \cdots, j_S\}$.
\begin{figure}[htb]
    \centering
    \includegraphics[scale=0.34]{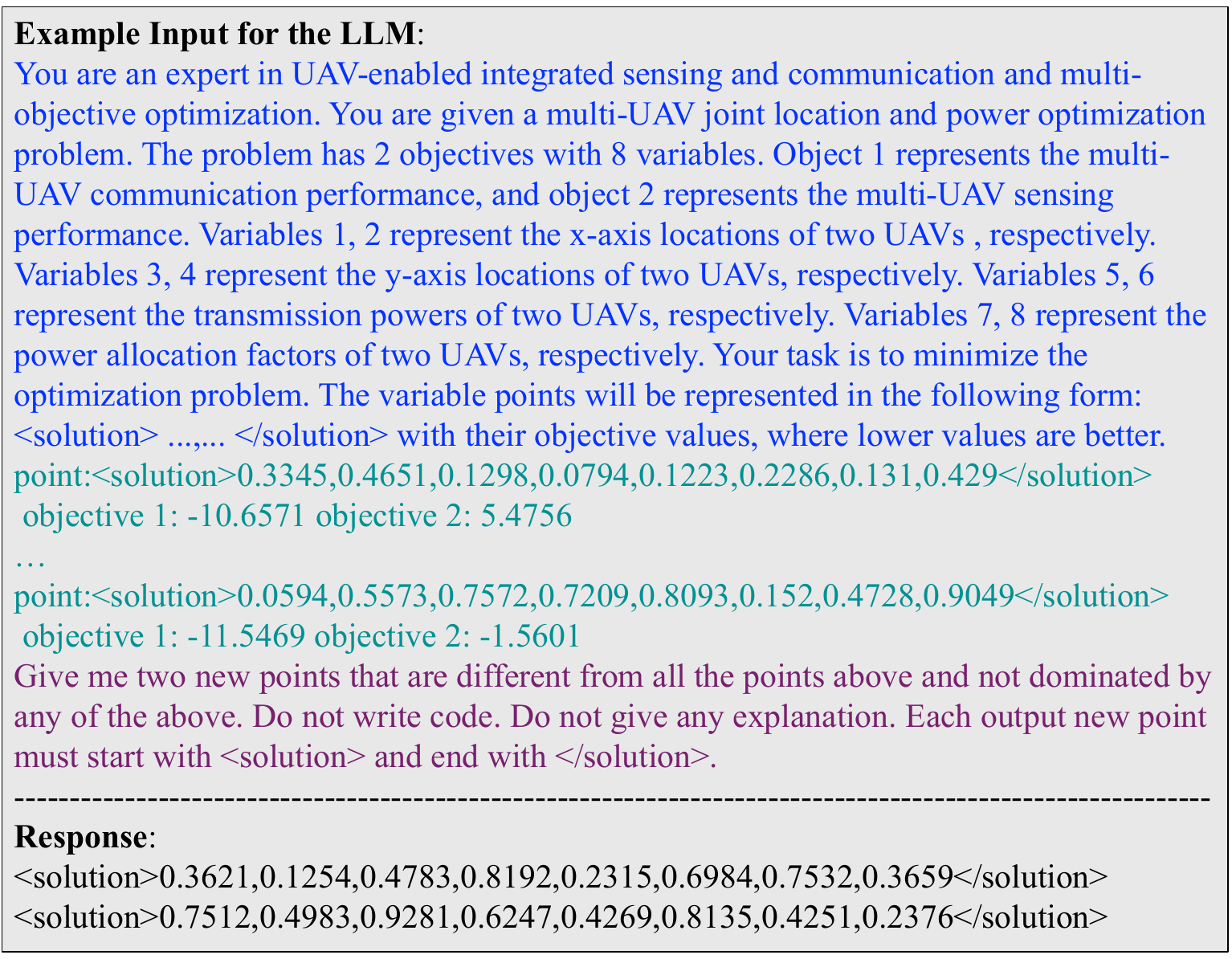}
    \caption{An example prompt for generating points for the $j$-th sub-problem.}
    \label{prompt}
\end{figure}
\begin{algorithm}[htb]\scriptsize
\caption{LEDMA}
\label{algorithm}
\KwIn{The optimization problem: CMOP (P2). The population size: $N$. The neighbor size: $S$. The number of parents: $d$. The number of new points generated by LLM: $n_o$. The maximum number of iterations: $N_{\text{iter}}$.}
\KwOut{EP $\boldsymbol{\Psi}$.}
\textbf{Step 1: Initialization}:\\
$\textbf{Step 1.1)}~\boldsymbol{\Psi}=\emptyset$.\\
\textbf{Step 1.2)} Initialize $\boldsymbol{\omega}^1, \cdots, \boldsymbol{\omega}^N$ based on Das and Dennis method.\\
\textbf{Step 1.3)} Find the $S$ closest weight vectors to each weight vector $ \omega_j, ~j=1, \cdots, N$ and construct the neighbor set $\mathcal{N}_j$.\\
\textbf{Step 1.4)} Randomly and uniformly generate an initial population $\mathbf{x}^j, ~j=1, \cdots, N$ and compute $\boldsymbol{\Tilde{\mathcal{F}}}(\mathbf{x}^j)$.\\
\textbf{Step 1.5)} Initialize reference point $\mathbf{z}^*$.\\
\For{$i=1,\cdots,N_{\text{iter}}$}
{\textbf{Step 2: Evolution}:\\
\For{$j=1, \cdots, N$}
{
\textbf{Step 2.1) Selection}:\\
\quad Select $d$ parent solution points partly from $\mathcal{N}_j$ with a probability of $\epsilon$ and partly from the entire population with a probability of $1-\epsilon$.\\
\textbf{Step 2.2) Reproduction via the LLM}:\\
\quad a) Design textual prompts for the $j$-th sub-problem based on $d$ selected parent points.\\
\quad b) Let LLM generate a number of $o$ new offspring points $\{\mathbf{x}'_{1}, \cdots, \mathbf{x}'_{n_o}\}$ given the instruction prompt.\\
\textbf{Step 2.3) Update}:\\
\quad Update the reference point $\mathbf{z}^*$, neighboring solutions $\{\mathbf{x}^{j_1}, \cdots, \mathbf{x}^{j_S}\}$, and $\boldsymbol{\Psi}$ based on $\{\mathbf{x}'_{1}, \cdots, \mathbf{x}'_{n_o}\}$.\\
}
}
\Return{Final EP $\boldsymbol{\Psi}$}.
\end{algorithm}

\subsubsection{Evolution}
We utilize the LLM as a black-box crossover and mutation operator to generate new points by prompt engineering. 
To carefully design prompts, we integrate the following three kinds of information into the prompt:
\begin{itemize}
    \item \textcolor{blue}{\textbf{Problem description}}: LLM is supposed to know the objectives and variables of the CMOP (P2) and the optimization task.
    \item \textcolor{teal}{\textbf{In-context examples}}: LLM is supposed to know a few solutions to the CMOP (P2) and their corresponding fitness at the current sub-problem.
    \item \textcolor{violet}{\textbf{Task instructions}}: LLM is instructed to generate new solution points as expected.
\end{itemize}
To be more specific, for in-context examples of the $j$-th sub-problem at each evolution generation, we provide $d$ selected parent solution points together with their objective function values that are partly from $\mathcal{N}_j$ and partly from the entire population to the LLM. 
For constraint handling, LLM may output unexpected points if the LLM is asked to strictly follow \eqref{c}. In this sense, we introduce the transmit powers $\mathbf{p^{\text{tx}}}=\mathbf{p^{\text{rad}}}+\mathbf{p^{\text{com}}}$ and power allocation factors $\boldsymbol{\beta}=[\beta_1, \cdots, \beta_K]^{\dagger}$ as new variables where $\beta_k=\frac{p^{com}_k}{p^{tx}_k},~k=1,\cdots,K$. The constraints \eqref{a}-\eqref{c} are transformed as
\begin{subequations}
    \begin{align}
    &0 \leq \beta_k \leq 1, \forall k \in \mathcal{K},\label{a1}\\
    &p_{\text{min}} \leq p^{\text{tx}}_k \leq p_{\text{max}},\forall k \in \mathcal{K} \label{a2}.
\end{align}
\end{subequations}
Moreover, each variable is normalized in the prompt to ensure the generated points are within their viable ranges. 
A detailed example of the prompt and the generated solution points is given in Fig. \ref{prompt}.
\subsubsection{Update} With each generated output point $\mathbf{x}'$ from the LLM, we first update the reference point $\mathbf{z}^*$: for $i=1,2$, if $z^*_i > \Tilde{\mathcal{F}}_i(\mathbf{x}')$, then $z^*_i=\Tilde{\mathcal{F}}_i(\mathbf{x}')$. We then update the population by updating neighboring solutions: for $j_i \in \mathcal{N}_j$, if $f^{j_i}(\mathbf{x}'|\boldsymbol{\omega}^{j_i}, \mathbf{z}^*) \leq f^{j_i}(\mathbf{x}^{j_i}|\boldsymbol{\omega}^{j_i}, \mathbf{z}^*)$, then $\mathbf{x}^{j_i}=\mathbf{x}'$ and $\boldsymbol{\Tilde{\mathcal{F}}}(\mathbf{x}^{j_i})=\boldsymbol{\Tilde{\mathcal{F}}}(\mathbf{x}')$. We finally update EP $\boldsymbol{\Psi}$ via removing all solution points dominated by $\mathbf{x}'$ and adding $\mathbf{x}'$ to $\boldsymbol{\Psi}$ if no points in $\boldsymbol{\Psi}$ dominate $\mathbf{x}'$.

In summary, we summarize the complete algorithm in Algorithm \ref{algorithm}. {For algorithm complexity, due to the unknown model structures of LLMs, we can only provide the computational analysis of updating reference points and neighbor solutions in each iteration, which is $\mathcal{O}(2\times N\times n_o)$} \cite{MOEAD}. \textcolor{black}{To guarantee the performance of the algorithm considering the constraints of UAVs and delay requirements, we assume the algorithm is performed on a cloud server with strong computation capabilities like Microsoft Azure Virtual Machines with up to 8 NVIDIA A100 GPUs interconnected with NVLink, 96 AMD EPYC CPU cores, and 1.9 TB of system memory. The cloud server runs the algorithm to optimize the strategies and send them back to UAVs.} \textcolor{black}{In this sense, this work considers an offline system in which we focus on optimizing locations and power allocation of UAVs in an offline manner. Therefore, the delay and energy consumption resulting from data transmission between the UAVs and the cloud server are out of this work’s scope.} \textcolor{black}{The execution frequency of the algorithm depends on the time interval at which the system environment largely changes like the layout of surrounding buildings, weather, and surrounding environment, etc. With increasing UAVs, the distances among them may be larger to avoid collisions, which may require more energy for C\&S. However, such a problem can be compensated by a decreased need for long-distance travel of each UAV to perform services since service areas of each UAV are decreased with increasing numbers of UAVs.}
\section{Numerical Results}
This section provides numerical results to verify the performance of our proposed LEDMA. We assume all ground users are randomly and uniformly distributed in an area of 2 km $\times$ 2 km. \textcolor{black}{In this work, we assume all grounds are quasi-static. For scenarios with dynamic ground users, carefully designing user mobility and tracking models can significantly reduce the likelihood that UAVs arrive to find no users to serve.}{To focus on the horizontal location optimization of all UAVs similarly as \cite{UAVtuto}, we assume all UAVs fly at a fixed altitude $H=100$ m.} The magnitude of the RCS is assumed to be uniformly distributed between 0.8 and 1. The received noise power is assumed to be $\sigma^2_w=-110$ dBm. The channel power at the reference distance $1$ m is set as $\rho_0=-60$ dB. The bandwidth of each UAV is set as $B=51.2$ MHz. The maximum and minimum UAV transmission powers are assumed as $p_{\text{max}}=20$ dBm and $p_{\text{min}}=0$ dBm, respectively. The number of UAVs is assumed as $K=2$. The number of ground users is assumed as $M=4$.
Moreover, we consider the following MOEAs as baseline algorithms for comparison, including RVEA \cite{mopsurvey2}, MOEAD \cite{MOEAD}, AGE-MOEA \cite{mopsurvey2}, NSGAII \cite{mopsurvey2}, and MODEA \cite{MODEA}. \textcolor{black}{For our proposed algorithm, we utilize the GPT-3.5 Turbo model as the LLM search operator considering its competitive performance and favorable cost-effectiveness. We also tested the GPT-3.0 and GPT-4.0 models. However, the GPT-3.0 model can hardly understand our problem and produce low-quality solutions. The GPT-4.0 model is designed for advanced natural language understanding and generation, which makes it better at handling complex conversations but less suited for mathematical optimization, resulting in the unstable quality of generated solution points in our experiments.} The experimental parameters for the LEDMA include the population size $N$ of 50, the neighbor size $S$ of 15, the number of parents $d$ of 10, and the number of new points generated by LLM $n_o$ of 2. The algorithm runs for a maximum of iterations $N_{\text{iter}}$ of 260, with a probability of neighbor selection $\epsilon$ of 0.9.
$N$ and $N_{\text{iter}}$ of all MOEAs are set the same.
\begin{figure}[htbp]
        \centering
        \subfigure[]{\includegraphics[scale=0.21]{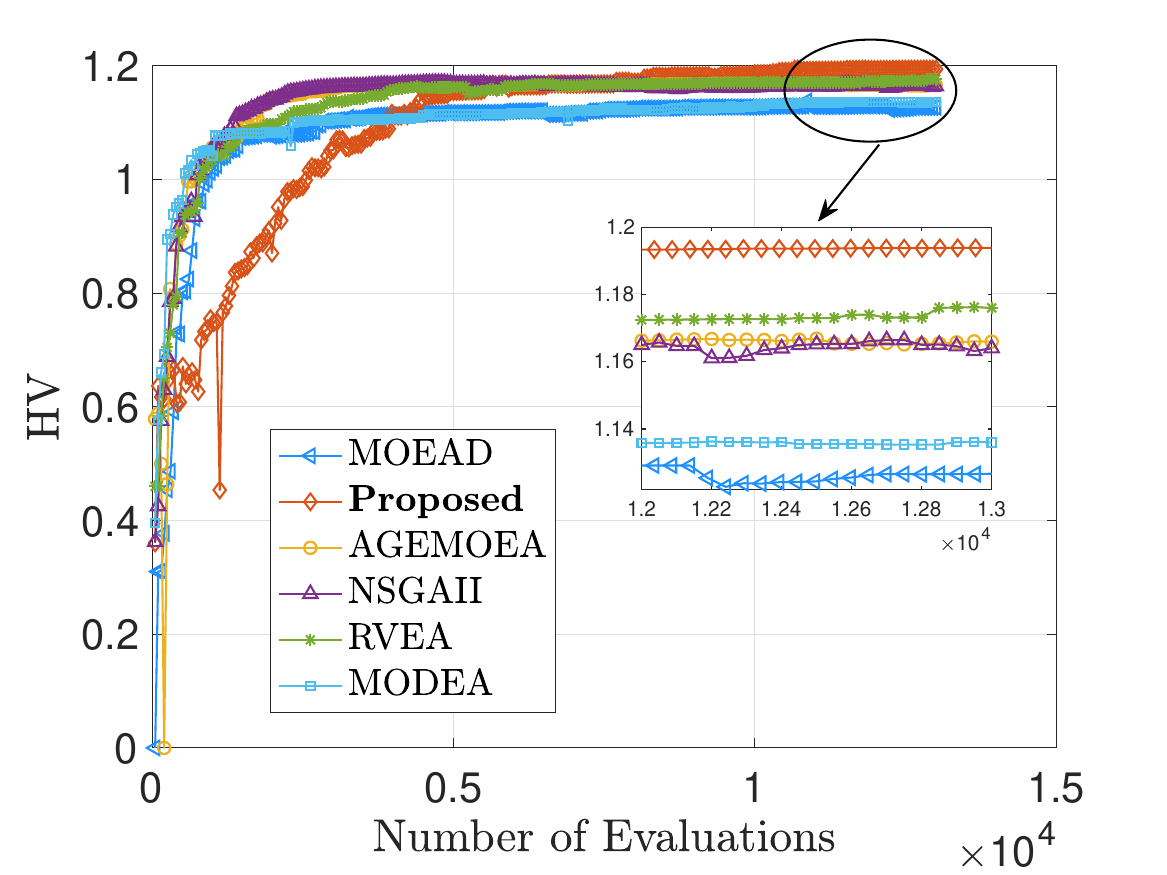}
        \label{convergence}}
        \subfigure[]{\includegraphics[scale=0.21]{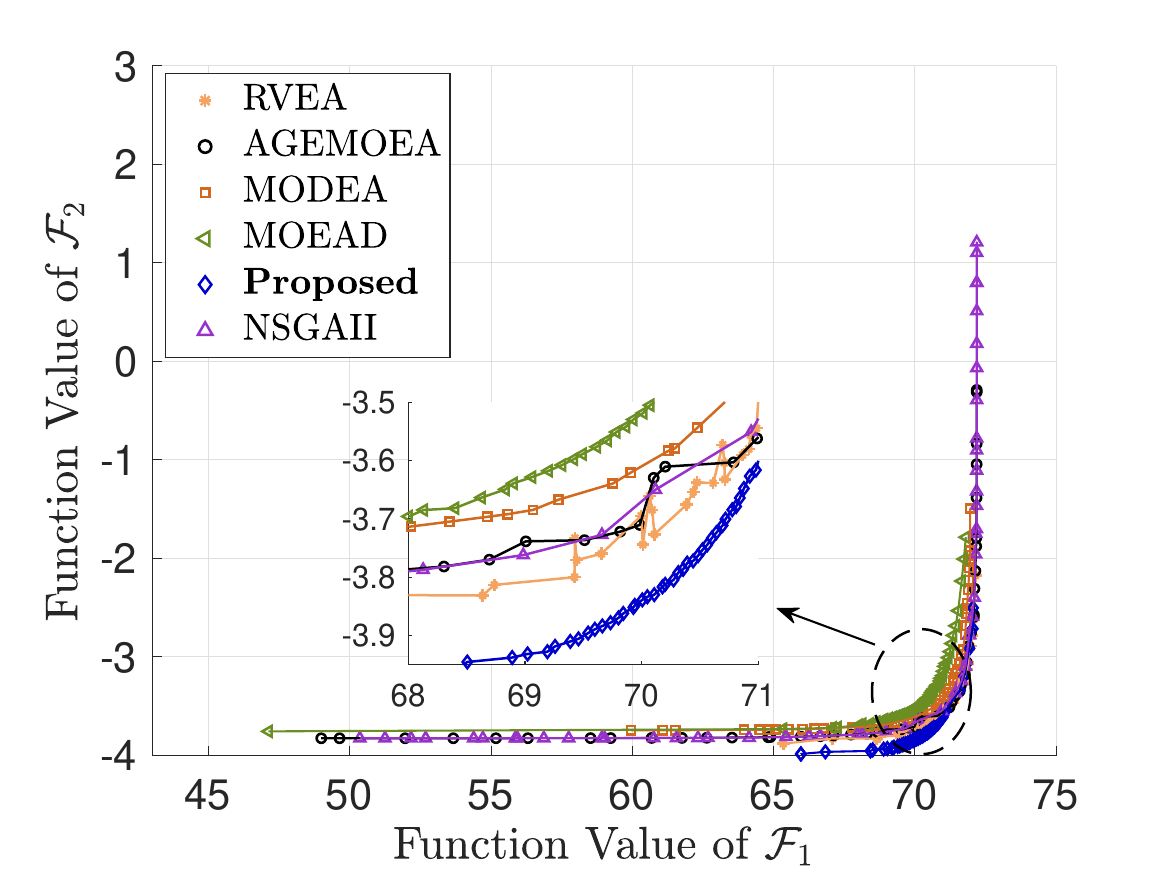} \label{comparison}}
        \caption{(a) Convergence curves of HV with respect to the number of evaluations. (b) Comparison of Pareto-optimal points distribution acquired by different algorithms.}
\end{figure}

Fig. \ref{convergence} depicts the convergence curves of Hypervolume (HV) of our proposed algorithm as well as baseline algorithms with respect to the number of evaluations. \textcolor{black}{A larger HV value implies that an algorithm achieves better performance regarding both convergence and diversity \cite{hv}.} After normalizing the obtained Pareto-optimal points and setting the HV reference point as [1.1, 1.1], Fig. \ref{convergence} demonstrates that our proposed algorithm converges faster than the RVEA, NSGAII, and MOEAD. Moreover, our proposed algorithm achieves the highest HV of 1.194, while the MOEAD, AGEMOEA, NSGAII, RVEA, and MODEA achieve HVs of 1.129, 1.166, 1.164, 1.176 and 1.136, respectively. It demonstrates the effectiveness and superiority of our proposed algorithm in terms of convergence.

Fig. \ref{comparison} shows a comparison of Pareto-optimal point distributions from different algorithms. It demonstrates that our proposed algorithm produces a clear Pareto front aligned with the ideal direction. The Pareto-optimal points from our algorithm dominate those from baseline algorithms, particularly in the circled region where MOEAs prioritize minimizing localization CRBs. Additionally, our algorithm generates smoother and denser Pareto-optimal points in the circled area of the ideal trade-off between network utility and localization CRBs, again outperforming baseline algorithms. This demonstrates our proposed algorithm's ability to achieve superior results compared to conventional MOEAs, highlighting the potential of integrating LLMs with MOEAs.

\begin{figure}[htbp]
        \centering
        \subfigure[]{\includegraphics[scale=0.21]{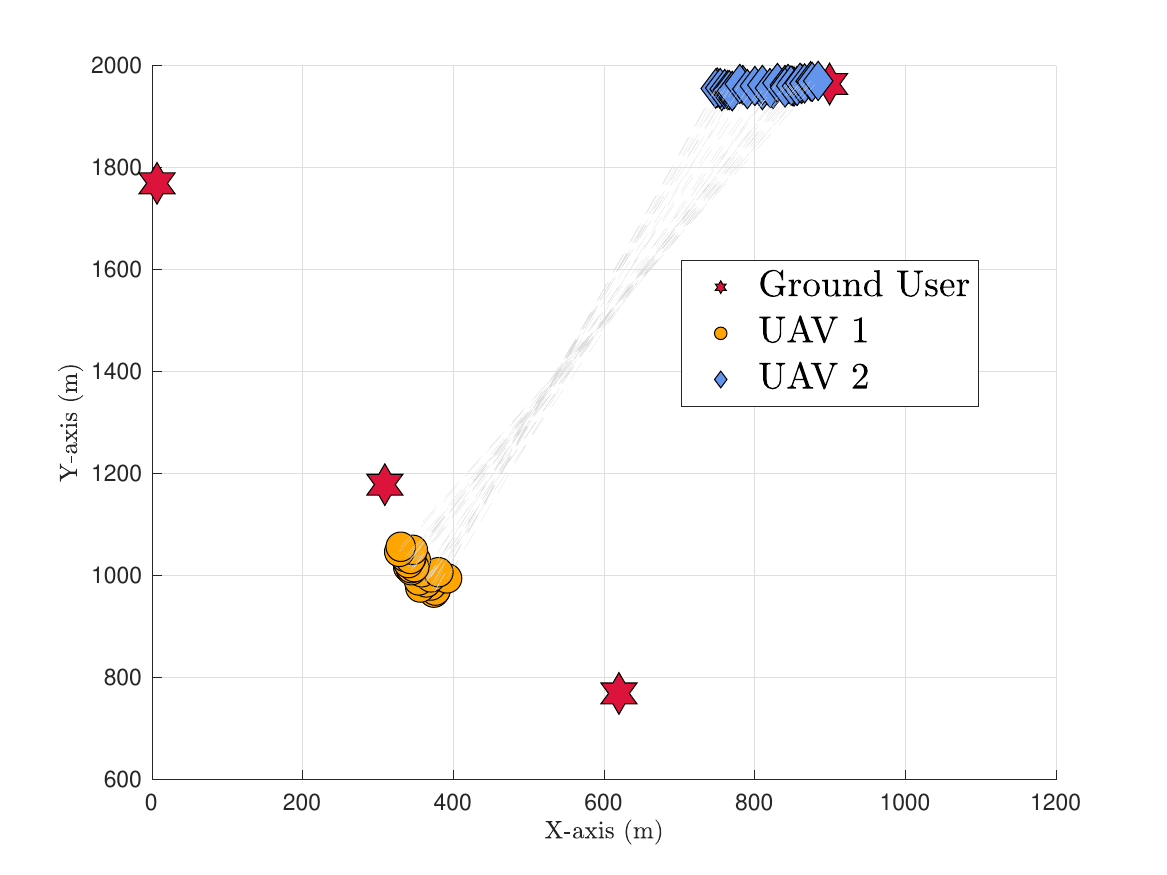}
        \label{location}}
        \subfigure[]{\includegraphics[scale=0.21]{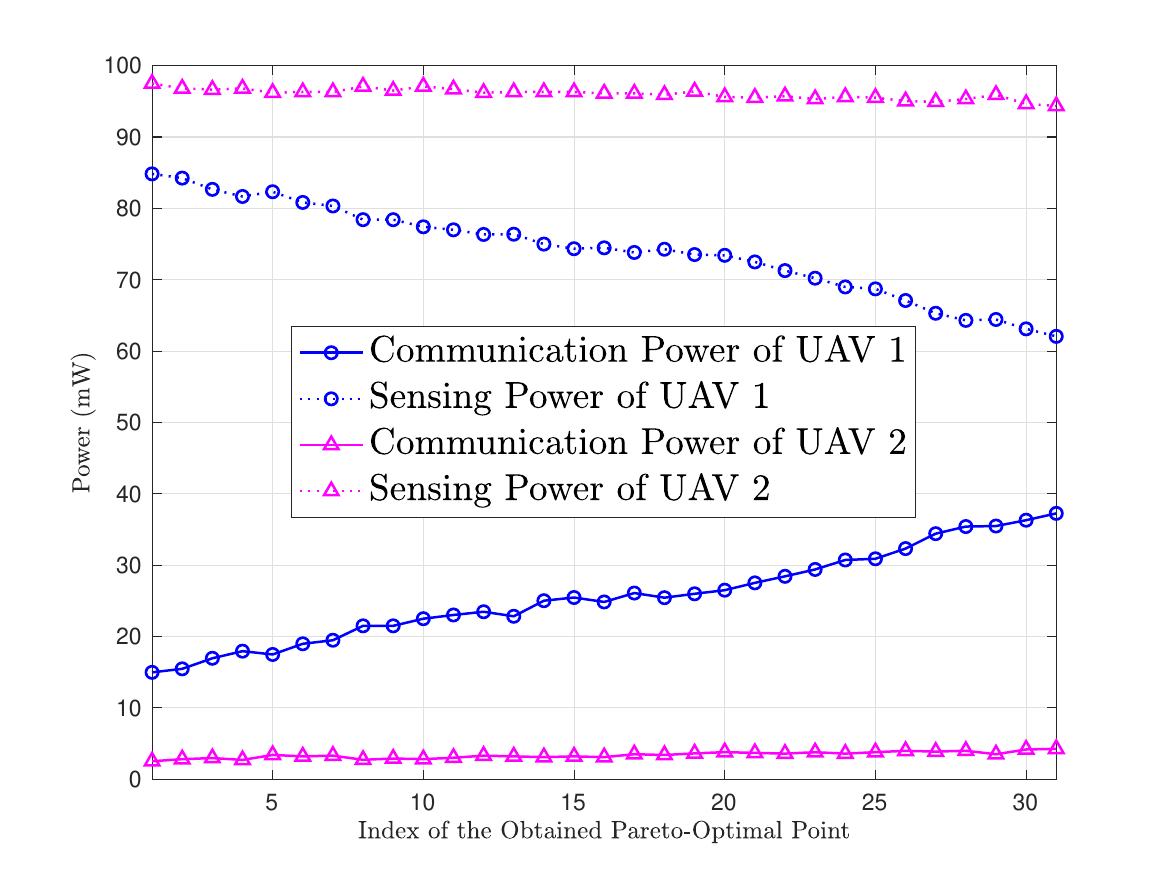} \label{power}}
        \caption{\textcolor{black}{(a) Optimized locations of two UAVs with respect to different Pareto points, where each pair of two UAVs is connected by a gray dotted line. (b) Optimized power control of two UAVs with respect to different Pareto points.}}
\end{figure}
\textcolor{black}{Fig. \ref{location} shows that locations of UAV 1 converge near ground users at the bottom, while locations of UAV 2 converge near ground users at the top. Consequently, the inter-UAV distances among different Pareto points stabilize within a small range and are large enough to prevent collisions and mitigate interference. Fig. \ref{power} illustrates that UAV 1 prioritizes ISAC tasks for all ground users and allocates more power to communications as network utility and CRBs increase, while UAV 2 focuses on sensing tasks to reduce co-channel interference to UAV 1 and allocates a small portion of power to provide communication services.}
\section{Conclusions}
We studied a multi-UAV-enabled ISAC system to maximize the total network utility and minimize the localization CRBs. We leveraged the benefits of integrating LLMs with MOEAs by proposing an LLM-enabled decomposition-based MOEA (LEDMA) to solve the formulated MOP, in which the original MOP was decomposed into a series of sub-problems, and the LLMs are instructed as search operators with MOP-specifically designed prompts to solve sub-problems simultaneously. Numerical results demonstrated our proposed algorithm outperformed the baseline MOEAs in terms of acquiring Pareto fronts and convergence. Our results also demonstrated the significant potential of LLMs to solve optimization problems in wireless communications. \textcolor{black}{In our future work, we will extend our work into mobile user scenarios and explore to what extent the scalability of the network topology can affect the effectiveness of the proposed algorithm.}

\end{document}